\documentstyle[12pt]{article}

\catcode`\@=11

\begin{document}

\baselineskip 24pt
\newcommand{\numero}{YCTP - P12-94}
\newcommand{\hep}{hep-ph/9410246}
%Enter SHEP preprint number
\newcommand{\titre}{DYNAMICAL GENERATION OF THE CKM MATRIX}
\newcommand{\titreb}{}
\newcommand{\auteura}{Nick Evans}
\newcommand{\addressa}{ }
\newcommand{\auteurc}{D.A. Ross }
\newcommand{\beq}{\begin{equation}}
\newcommand{\eeq}{\end{equation}}
\newcommand{\Fn}{\mbox{$F(p^2,\Sigma)$}}

\newcommand{\addressc}{Yale University   \\ Sloane Physics
Laboratory, \\ P.O.
Box 208120, \\ New Haven,\\ Connecticut 06520-8120 \\ U.S.A }
\newcommand{\abstrait}{We argue that in dynamical models of fermion
masses,
which explain the three mass scales of the generations of fermions
with three
separate heavy scales above the electroweak symmetry breaking scale,
the off
diagonal CKM matrix elements must be generated at a fourth scale. The
simplest
form for the additional mass generation is a mechanism that has an
$SU(3)_L
\otimes SU(3)_R$ family symmetry. We show that such an ansatz can
produce the
up down mass inversion and the observed Cabibbo angle. If the mass
terms for
the third family quarks are enhanced by additional interactions the
full CKM
matrix may be realized. We present a toy  model which realizes vacua
with these
masses.}

\begin{titlepage}
\hfill \numero

\hfill \hep
\vspace{.2in}
\begin{center}
{\large{\bf \titre }}
{\large{\bf \titreb}}
\bigskip \\by\bigskip\\ \auteura \bigskip \\ \addressc \\

\renewcommand{\thefootnote}{ }
\vspace{.7 in}
{\bf Abstract}
\end{center}
\abstrait
\bigskip \\
\end{titlepage}

\def\id{\rlap{1}\hspace{0.15em}1}

Recent work has shown that one family extended technicolour (ETC)
models
\cite{ETC} may be built that are both plausibly consistent with the
precision
data from LEP \cite{Realistic} and compatible with the third family
masses
\cite{Mass}. These models have a techni-family mass spectrum of the
form

\beq M_U = M_D \sim 400GeV, \hspace{1cm} M_E \sim 150GeV,
\hspace{1cm} M_N \sim
50GeV \eeq

\noindent In Ref\cite{Mass} the author has argued that if there is a
single
feed down ETC interation for each of the first and second families
then those
families' masses (neglecting neutrinos) follow naturally from this
spectrum and
the large top mass (generated by a direct, but sub-critical, top
condensating
ETC interaction). The analysis of Ref\cite{Mass} does not however
generate the
up down mass inversion or the off diagonal CKM matrix elements. With
the
notable exception of Refs\cite{Moose,Georgi,King} there has been
little
discusion of the CKM matrix generation in dynamical models in the
literature.
In this letter we propose a simple form for the off diagonal mass
terms and
show that it is compatible with the observed CKM matrix elements. In
addition
the ansatz is simultaneously capable of generating the up down mass
inversion.
Finally we present a toy model with vacua that generate these mass
matrices. A
naive analysis of the effective potential in the model does not
however favour
these precise vacua.

ETC models generate the light fermion masses by the exchange of heavy
gauge
bosons which couple the light fermions to techni-fermion condensates
with
coupling $g^2/M^2_{ETC}$. The hierarchy of three families may either
be
generated by three separate magnitude couplings at one ETC mass scale
or by
three ETC mass scales. Though the former possibility may be realized
in models
with complicated mixings and/or many ETC gauge groups (eg
Ref\cite{Moose}) we
consider the latter with a single ETC group for the entire
techni-family more
natural. However, if the ETC breaking is of the form

\beq SU(N+3)_{ETC} {\Lambda_1 \atop \longrightarrow} SU(N+2)
{\Lambda_2 \atop
\longrightarrow} SU(N+1) {\Lambda_3 \atop\longrightarrow }SU(N)_{TC}
\eeq

\noindent then the gauge eigenstates corresponding to the light three
families
are strictly defined by the breaking pattern. The mass eigenstates
are
identical to the gauge eigenstates and the model can not give rise to
relative
rotations between the up and down type quark mass matrices.
Additional dynamics
at a fourth (or more) scale is required to generate the CKM matrix
elements.

In order to show how off diagonal fermion mass terms may be generated
let us
first review the vacuum structure of QCD. We shall just consider the
QCD
interactions of the up and down quarks. The theory has an $SU(2)_L
\otimes
SU(2)_R$ flavour symmetry that is broken by the current quark masses

\beq M_c = \left( \begin{array}{cc} m_u & 0 \\ 0 & m_d \end{array}
\right) \eeq

\noindent This breaking term allows us to distinguish a number of
distinct
vacua described by the condensates

\beq \left( \bar{u_L} , \bar{d_L} \right)  U \left( \begin{array}{c}
u_R \\ d_R
\end{array} \right)  \eeq

\noindent where  $U$ is an element of SU(2). Clearly except when U is
the
identity these vacua in addition to the current masses give off
diagonal
dynamical masses. In QCD the current masses perturb the energy of
each of these
vacua through the contributions to the effective potential  of the
form

\beq - Tr (M_c^{\dagger} U)  + h.c. \eeq

\noindent generated by the diagram in Fig1. The lowest energy vacuum
is that
with U equal to the identity (which maximises the trace) and no off
diagonal
terms are observed. This vacuum is also prefered since it leaves the
U(1) of
QED unbroken.

We propose that a similar vacuum structure could give rise to the CKM
matrix
mixings in technicolour models though the potential would have to
take a
different form.

\noindent {\bf  An Ansatz With SU(3) Family Symmetry}

The simplest assumption about the dynamics responsible for the CKM
elements is
that it is $SU(3)_L \otimes SU(3)_R$ family symmetric and flavour
blind. Our
ansatz for the additional contribution to the up and down type quark
mass
matrices are thus proportional to the identity in some basis rotated
by an
element of SU(3) relative to the usual ETC interactions in Eqn(2). We
have

\beq \begin{array}{ccc}
M_U & =& \left( \begin{array}{ccc} m_t^{ETC} & 0 & 0\\
                                                                 0 &
m_c^{ETC}
& 0\\
                                                                 0 &
0 &
m_u^{ETC}  \end{array} \right) +
                   \tilde{m}  {\huge U} \\
&&\\
M_D & =& \left( \begin{array}{ccc} m_b^{ETC} & 0 & 0\\
                                                                 0 &
m_s^{ETC}
& 0\\
                                                                 0 &
0 &
m_d^{ETC}  \end{array} \right) +
                   \tilde{m}  {\huge U} \end{array}  \eeq

\noindent where $\tilde{m}$ is the mass scale generated by the new
dynamics,
$m_i^{ETC}$ is the diagonal mass generated by the usual ETC
interactions and U
is an element of  SU(3). Following Ref\cite{Georgi} we parameterize U
in terms
of an SU(2) matrix, $\Sigma$, and phases

\beq  U = \left( \begin{array}{cc}
e^{-i\phi}(1-(1-\cos\theta)uu^{\dagger})\Sigma &  -\sin\theta u\\
                                                           \sin\theta
u^{\dagger} \Sigma  &   \cos\theta e^{i\phi}
                             \end{array} \right) \eeq

\noindent with $u$ a unimodular 2-vector

\beq u = \left( \begin{array}{c}   a e^{i\alpha}  \\
(1-a^2)^{\frac{1}{2}}
e^{i\beta}  \end{array} \right) \eeq

The diagonal mass matrices are given by

\beq \begin{array}{ccc}  M_U^{diag} & = &  L_{U}^{\dagger}  M_U R_{U}
\\
&&\\
M_D^{diag} & = & L_{D}^{\dagger} M_D R_{D} \end{array} \eeq

\noindent where $L_{i}$ and $R_{i}$ are left and right handed SU(3)
family
transformations that diagonalize the hermitian matrices
$M_i^{\dagger}M_i$ and
$M_iM_i^{\dagger}$ respectively. The CKM matrix is then given by

\beq K = L_{D}^{\dagger} L_{U}  \eeq

\noindent {\bf The Cabibbo Sector}

For simplicity we first consider numerical results for only the
lightest two
families of quarks. The CKM matrix phases may then be rotated away by
global
family transformations. Standard ETC interactions do not naturally
generate the
up down mass inversion (prefering in fact  the inverse) or CKM
mixings. We
shall therefore take as our ansatz

\beq   M_U = \left( \begin{array}{cc}  1.5 & 0 \\ 0 & 0.005
\end{array} \right)
+ \tilde{m} U \hspace{.3cm} GeV \eeq

\beq   M_D = \left( \begin{array}{cc}  0.2 & 0 \\ 0 & 0.002
\end{array} \right)
+ \tilde{m} U \hspace{.3cm} GeV \eeq

\noindent Mixing angles will only result from off diagonal mass terms
(on
diagonal masses simply correspond to rescaling the ETC feed down
masses) and
therefore we take as our  SU(2) matrix

\beq U = \left( \begin{array}{cc} 0 & 1 \\ 1&0 \end{array} \right)
\eeq

\noindent and search for values of $\tilde{m}$ compatible with the
observed
masses and mixings. We find for $\tilde{m} = 0.057GeV$

\beq \begin{array}{ll}
m_c = 1.5GeV & m_s = 0.2GeV \\
m_u  = 0.003 GeV & m_d = 0.013GeV \end{array}
|K| = \left( \begin{array}{cc}
        0.975 & 0.22\\
        0.22 & 0.975 \end{array} \right) \eeq

The ansatz correctly describes both the up down mass inversion and
the Cabibbo
sector of the CKM matrix.

\noindent {\bf Three Families}

In extending the ansatz to the third family we shall again
concentrate on the
CKM matrix magnitudes since the CKM matrix phase is ill determined. U
now has
three parameters ($a$,$\theta$ and its equivalent angle in $\Sigma$)
that
determine the magnitudes of the CKM elements and six phases.
Solutions again
exist for this ansatz that are compatible with the up and down quark
masses and
the Cabibbo angle. For example

\beq \begin{array}{ccc}
M_U & =& \left( \begin{array}{ccc} 170 & 0 & 0\\
                                                                 0 &
1.5 & 0\\
                                                                 0 &
0 & 0.005
\end{array} \right) +
                   0.052  {\huge U} \hspace{0.3cm} GeV\\
&&\\
M_D & =& \left( \begin{array}{ccc} 5& 0 & 0\\
                                                                 0 &
0.2& 0\\
                                                                 0 &
0 & 0.002
\end{array} \right) +
                   0.052  {\huge U} \hspace{0.3cm} GeV\end{array}
\eeq

\noindent where

\beq |U| = \left( \begin{array}{ccc} 0.76 & 0.42 & 0.63 \\
                                                              0.41 &
0.28 &
0.87 \\
                                                              0.48 &
0.86 &
0.04 \end{array} \right) \eeq

\noindent  diagonalizes to give

\beq \begin{array}{ll}
m_t = 170 GeV & m_b = 5GeV \\
m_c = 1.5GeV & m_s = 0.2GeV \\
m_u  = 0.007 GeV & m_d = 0.013GeV \end{array}
|K| = \left( \begin{array}{ccc}
        0.9999 & 0.006 & 0.003\\
        0.005  & 0.975 & 0.22\\
        0.004 & 0.22 & 0.975\end{array} \right) \eeq

\noindent  We note that these numerical results are not the result of
any fine
tuning in the ansatz's parameter space but are stable over a range of
parameters other than the  requirement that $|U_{33}|$ is at least an
order of
magnitude suppressed relative to the other elements of U. The third
family
mixing angles can be raised to their observed values by increasing
$\tilde{m}$
but at the expense of increasing the up and down quark masses above
their
experimental limits.

The third family mass elements may be enhanced relative to the other
families'
mass terms if the ETC model has strong extended technicolour
interactions.
Ref\cite{Mass} makes use of strong horizontal self interactions in
the techni
and third families to break the SU(8) flavour symmetry in the pattern
observed
in the light fermion masses. The simplest realization of this  global
symmetry
breaking would be through the addition of an extra strong (but not
super-critical) broken U(1) gauge interaction with appropriate
couplings. If
such a gauge boson mixed with the standard ETC gauge boson associated
with the
diagonal generator in the breaking $SU(N+1)_{ETC} \rightarrow
SU(N)_{TC}$ then
the self coupling of the third generation would naturally be greater
than that
of the first and second generations. It is then plausible that the
third family
quark mass elements may be enhanced relative to those of the lighter
two
families. Parameter ranges then exist that reproduce the full CKM
matrix, for
example with U in Eqn(15)

\beq |U| = \left( \begin{array}{ccc} 0.92 & \zeta 0.38 & \zeta 0.09
\\
                                                              \zeta
0.07 & 0.07
& 0.99 \\
                                                              \zeta
0.34 & 0.92
& 0.04 \end{array} \right) \eeq

\noindent and with $\zeta = 10$ (note that when when $\zeta =1$ U is
unitary
and that a parameter multiplying the on diagonal third family mass
terms simply
corresponds to a rescaling of the ETC masses in Eqn(15)) we obtain

\beq \begin{array}{ll}
m_t = 170 GeV & m_b = 5GeV \\
m_c = 1.5GeV & m_s = 0.2GeV \\
m_u  = 0.007 GeV & m_d = 0.014GeV \end{array}
|K| = \left( \begin{array}{ccc}
        0.9993 & 0.04 & 0.003\\
        0.04   & 0.975 & 0.22\\
        0.003 & 0.22 & 0.975\end{array} \right) \eeq

In a realistic model we might expect different $\zeta$s in the up and
down type
quark mass sectors (corresponding to the difference in top bottom
mass
splitting) however this would simply add new parameters to the
ansatz. We
therefore expect that solutions fitting the observed CKM matrix
element values
would still exist.
We note that this ansatz is not postdictive having the same number of
input and
output parameters. Nevertheless it is interesting that this simple
ansatz is
viable.

\noindent {\bf  A Toy Model}

In this section we introduce a simple model that realizes vacua with
the mass
matrices discussed above. The model has a full family of fermions
that
transform under the usual  ETC group $SU(N+3)_{ETC}$ plus some
assumed
additional dynamics that break the SU(8) flavour symmetry of the
fermion
family. The simplest additional dynamics would be, as discussed
above, simply
to add in a new U(1) gauge interaction with different charges for
each of the
fermion flavours. We shall assume that the standard ETC dynamics and
this
additional sector give rise to the diagonal contibutions to the
fermion masses
in Eqn(6) and concentrate our discussion on the generation of the
extra mass
contributions which are family symmetric.

In addition to these usual dynamics the flavours of fermions that mix
(including at least all the quarks) transform under some SU(M+1)
gauge group
which we shall call ultracolour. This additional gauge group breaks
at some
high scale, $\Lambda_U$, according to

\beq SU(M+1)_{UC} \rightarrow SU(M) \eeq

\noindent The broken singlet forms the usual quark and techni-quark
sectors.
Whether an additional unification with SU(3) colour occurs we leave
for future
model builders. The ultracolour group then becomes confining,
generating
dynamical masses for the  ultracoloured fermions which feed down to
the
singlets through the gauge bosons that acquired masses at the scale
$\Lambda_U$. The ultracoloured partners of the usual quarks  have an
$SU(3)_L
\otimes SU(3)_R$ family symmetry under the ultracolour interactions
and will
thus give rise to  vacua with mixings between the families. These
mixings will
feed down to the standard quark sector producing masses that mix the
families
as described in Eqn(6). This dynamics is entirely analogous to that
in QCD with
the ETC fed down masses playing the part of the current quark masses
in Eqn(3).
Presumably the effective potential will therefore take the same form
as that in
Eqn(5) (again generated by the diagram in Fig 1 but with ultra-quarks
in the
loop) and the vacua corresponding to the ansatz in Eqn(18)  would not
be the
prefered one. However, it is possible that dynamics associated with
the high
breaking scales may introduce additional parameters that perturb the
effective
potential resulting in the realistic vacua. It is not our intention
in this
letter to propose a phenomenologically consistent model but merely
provide an
existence proof  of models which potentially give rise to CKM mixing.
In this
spirit also we shall not be troubled by our need to reconcile the
extra
$M\times(N+3)$ ultracoloured quark doublets with measurements of the
precision
parameter S \cite{Realistic} that prefer to minimize the number of
strongly
interacting doublets.

In this letter we have proposed that a family symmetric contribution
to the
quark mass matrices may be generated in ETC models at an additional
scale to
the usual ETC breaking scales. Such an ansatz is capable of
explaining the up
down mass inversion simultaneously with the Cabibbo angle. If the
third family
quark mass terms are enhanced relative to those of the lighter two
families the
ansatz can describe the full CKM matrix structure . We have discussed
a toy
technicolour model that possesses vacua consistent with our ansatz
for the
quark masses. The model serves to highlight the problem that the
naive
effective potential of such a model does not favour the precise
physical
vacuum. Nevertheless we hope that this discussion sheds light on the
form of
the CKM matrix and will be helpful in future model building.
\vspace{1in}

{\bf Acknowledgements}

The author would like to thank Tom Appelquist and Steve Selipsky for
helpful
discussions. \vspace{3in}

\noindent {\bf Figure Caption}

\noindent Fig1: Contribution to the effective potential from (ultra)
quark
loops.

\end{document}